\documentclass[preprint2]{aastex}
\usepackage{amssymb}
\usepackage{amsmath}
\usepackage{graphicx}
\usepackage{txfonts}
\usepackage{hyperref}

\begin{document}
\title{Nonlinear evolution of short-wavelength torsional Alfv\'en waves}

\author{S.~V.~Shestov\altaffilmark{1,2}}
\affil{Solar-Terrestrial Centre of Excellence --- SIDC, Royal Observatory of Belgium, Avenue Circulaire 3, 1180 Brussels, Belgium}

\author{V.~M.~Nakariakov}
\affil{Centre for Fusion, Space and Astrophysics, Department of Physics, University of Warwick, Coventry CV4 7AL, UK,\\
St. Petersburg Branch, Special Astrophysical Observatory, Russian Academy of Sciences, 196140, St Petersburg, Russia}
\and 
\author{A.~S.~Ulyanov and A.~A.~Reva and S.~V.~Kuzin}
\affil{Lebedev Physical Institute,  Leninskii prospekt, 53, 119991, Moscow, Russia}
\altaffiltext{1}{Lebedev Physical Institute,  Leninskii prospekt, 53, 119991, Moscow, Russia}
\altaffiltext{2}{s.shestov@oma.be}
\date{Received ..; accepted ...}

\begin{abstract}
We analyse nonlinear evolution of torsional Alfv\'en waves in a straight magnetic flux tube filled in with a low-$\beta$
plasma, and surrounded with a plasma of lower density. Such magnetic tubes model, in particular, a segment of a coronal
loop or a polar plume. The wavelength is taken comparable to the tube radius. We perform numerical simulation of the
wave propagation using ideal magnetohydrodynamics. We find that a torsional wave nonlinearly induces three kinds of
compressive flows: the parallel flow at the Alfv\'en speed, that constitutes a bulk plasma motion along the magnetic
field, the tube wave, and also transverse flows in the radial direction, associated with sausage fast magnetoacoustic
modes. In addition, nonlinear torsional wave steepens and its propagation speed increases. The latter effect leads to
the progressive distortion of torsional wave front, i.e. nonlinear phase mixing. Because of the intrinsic non-uniformity
of the torsional wave amplitude across the tube radius, the nonlinear effects are more pronounced in regions with higher
wave amplitude. They are always absent at the axes of the flux tube. In the case of a linear radial profile of the wave
amplitude, the nonlinear effects are localised in an annulus region near the tube boundary. Thus, the parallel
compressive flows driven by torsional Alfv\'en waves in the solar and stellar coronae, are essentially non-uniform in
the perpendicular direction. The presence of additional sinks for the wave energy reduces the efficiency of the
nonlinear parallel cascade in torsional Alfv\'en waves.  
\end{abstract}


\section{Introduction}
The existence of plane Alfv\'en waves was theoretically predicted by \citet{1942Natur.150..405A}, and since then the ubiquitous presence of Alfv\'en waves was found in magnetospheric, space, cosmic and laboratory plasmas \citep[e.g.][]{1995PhST...60...20U, Cheng198521, 1402-4896-2000-T84-014}. In solar physics Alfv\'en waves are mainly considered as a candidate for coronal heating due to their ability to freely propagate from lower layers of the solar atmosphere to the corona \citep[e.g.][]{1999ApJ...521..851R, 2017NatSR...743147S}. For example, \citet{2008A&A...478..921C} suggested that propagating Alfv\'en waves could locally heat coronal plasma threads. In addition, \citet{2008ApJ...675.1645F} proposed that a flare-generated large-scale torsional wave could accelerate electrons to high energies. Alfv\'en waves may also contribute to the acceleration of the  solar and stellar winds \citep[e.g.][]{1995ApJ...454..901C, 2009LRSP....6....3C, 2012ApJ...749....8M}, and collimate astrophysical jets \citep{2007MNRAS.376..457B}. 

In the presence of field-aligned plasma non-uniformities, typical for the corona, Alfv\'en waves appear in a form of torsional waves \citep[e.g.][]{2008ApJ...676L..73V}. Torsional Alfv\'en waves are azimuthal (rotational) perturbations of the plasma velocity accompanied by azimuthal components of the magnetic field. Linear torsional waves propagate along the magnetic field at the local Alfv\'en speed. 

The key element of the nonlinear evolution of linearly and elliptically polarised Alfv\'en waves is the ponderomotive force that is associated with the variation of the absolute value of the magnetic field in the wave \citep[e.g.][]{1971JGR....76.5155H, 1995PhPl....2..501T}. This variation results in the gradient of the magnetic pressure that induces plasma flows and hence changes the density of the plasma. The induced compressive perturbations have double the frequency of the mother Alfv\'en wave. The induced variations of the absolute value of the magnetic field and plasma density change the local values of the Alfv\'en speed, and cause the Alfv\'en wave self-interaction resulting in steepening of the wave front \citep[e.g.][]{1995JGR...10023413O, Zheng2016}. 

Nonlinear dynamics of Alfv\'en waves is actively studied analytically and numerically, with the main emphasis put on the study of plane waves. It has been shown that weakly-nonlinear plane Alfv\'en waves are governed by the Cohen--Kulsrud evolutionary equation that is a modification of the well-known Burgers equation on accounting for the cubic nonlinearity intrinsic to Alfv\'en waves \citep[e.g.][]{1974PhFl...17.2215C}. In addition, when the wave front is non-plane or oblique, Alfv\'en waves induce oblique compressive perturbations \citep[e.g.][]{1996ApJ...459..347M, Nakariakov1997}. The ponderomotive acceleration effects have been intensively studied in the context of acceleration of the solar and stellar winds \citep[e.g.][]{1998JGR...10323677O, 2000A&A...353..741N, 2011SSRv..158..339S}, and also of the first ionisation potential effect \citep[e.g.][]{2015LRSP...12....2L}.

Nonlinear evolution of plane Alfv\'en waves is strongly affected by transverse structuring of the plasma in the Alfv\'en speed and/or field-aligned plasma flows. The transverse structuring of the medium leads to the wavefront distortion, the wave becomes progressively non-planar, and the effect of phase mixing comes into play. It causes enhanced dissipation of Alfv\'en waves \citep[e.g.][]{Heyvaerts1983, 2002RSPSA.458.2307W}, and induces compressive perturbations that produce fast and slow magnetoacoustic waves \citep[e.g.][]{Nakariakov1997, 1998A&A...332..795N, Botha2000, 2002A&A...393..321T, 2003A&A...400.1051T}. 

However, in a plane wave the perpendicular scale, i.e. the length of the wave front, should be much larger than the parallel wavelength, which is rarely fulfilled. For example, in the solar corona the generation of a plane Alfv\'en wave with a 10-min period and the Alfv\'en speed of 1~Mm\,s$^{-1}$ the wave driver at the base of the corona should be much larger than 600~Mm (i.e. much larger than the radius of the Sun).   As the solar corona is structured into magnetic flux tubes, represented, for example, by coronal loops, plumes and various filaments, coronal Alfv\'en waves are likely to appear in the form of torsional, rather than plane perturbations. Torsional waves have received attention in the context of heating of the plasma in coronal loops. In particular, \citet{1996A&A...306..610P, 1997A&A...320..305R} considered dissipation of finite amplitude waves driven periodically at one footpoint of a closed magnetic structure. On the other hand, \citet{2010ApJ...712..494A} established several constraints on the parametric range of Alfv\'en waves as a coronal heating mechanism.

The understanding of the appearance of coronal Alfv\'en waves in the torsional form stimulated intensive studies of nonlinear evolution of torsional waves. \citet{1999ApJ...514..493K} and 
\citet{VasheghaniFarahani2011} numerically and analytically, respectively, demonstrated that propagating torsional Alfv\'en waves, similarly to plane waves, induce compressive perturbations due to the ponderomotive force associated with nonlinear effects. \citet{VasheghaniFarahani2012} found that weakly-nonlinear self-interaction of torsional Alfv\'en waves results in the Cohen--Kulsrud steeping of the wave profile, while the coefficients are different from the case of a plane wave. The self-interaction occurs due to the nonlinear excitation of compressive waves and their back-reaction on the mother torsional waves. \citet{Fedun2011} showed by means of numerical simulations that chromospheric magnetic flux tubes can act as a frequency filter for torsional waves. \citet{2014SoPh..289.4069M} studied analytically the resonant generation of compressive waves by nonlinear coupling of two torsional waves propagating in the opposite directions. \citet{2015A&A...577A.126M, 2017SoPh..292...31W} highlighted the effect of the expanding magnetic tube on the evolution of torsional Alfv\'en waves, and emphasised the importance of the stratification. \citet{2016ApJ...817...92W} modelled the propagation of torsional waves in the presence of  slow compressive shocks. 

However, the efficiency of nonlinear interaction of torsional and compressive waves is not understood in detail so far. In particular, the role of the transverse profile of the torsional wave needs to be revealed. The main motivation for this study is the intrinsic non-uniformity of the wavefronts of torsional waves. Indeed, as the finite amplitude effects increase the wave propagation speed \citep[e.g.][]{1974PhFl...17.2215C}, and as the torsional wave has a zero amplitude at the axis of the guiding magnetic flux tube and increases to the tube boundary, the propagation speed is always non-uniform across the field. It creates the conditions for phase mixing even if the Alfv\'en speed inside the flux tube is uniform, as the torsional waves have different nonlinear increase in the propagation speed at different distance from the tube axis. This nonlinear phase mixing leads to the generation of compressive perturbations. The induced compressive perturbations may lead to the torsional wave self-interaction. Also, in contrast with the incompressive Alfv\'en waves,  the induced compressive perturbations can be detected with imaging telescopes, and are subject to enhanced damping by various dissipative mechanisms. 

The aim of this paper is to study by means of 3D magnetohydrodynamic (MHD) modelling the effects of nonlinear excitation of compressive perturbations by a torsional wave of finite amplitude. We consider the parallel wavelength to be not much larger than the radius of the magnetic flux tube, and thus account for the nonlinear phase mixing. This approach is a finite-wavelength generalisation of the analytical theory developed by \citet{VasheghaniFarahani2011} in the thin flux tube approximation that corresponds to the  long wavelength limit. We neglect the effect of the stratification of the atmosphere, as the typical radii of coronal field-aligned plasma non-uniformities and hence the considered wavelengths are about 1--10~Mm, while the typical scale heights of the stratification exceed 50~Mm. Thus, we consider evolution of torsional waves in a flux tube that is uniform along its axis.
The paper is organised as follows: 
in Section~\ref{numerical_setup} we describe our numerical setup, in Section~\ref{results} we discuss the results obtained, and finally give conclusions in Section~\ref{conclusions}. Appendix~\ref{thintube} illustrates the nonlinear excitation of compressive perturbations by a long-wavelength torsional waves in the thin flux tube approximation, and Appendix~\ref{Aphasemix} illustrates the generation of compressive waves due to the phase-mixing effect. 

\section{Numerical setup \& Initial conditions}\label{numerical_setup}
\subsection{MHD equations and normalisation}
The simulations were performed using the numerical code MPI-AMRVAC \citep{2014ApJS..214....4P}. The code applies the Eulerian approach to the solution of the resistive MHD equations,
\begin{equation}
	\frac{\partial \rho}{\partial t} + \nabla \cdot (\rho \mathbf{v}) = 0,
\end{equation}
\begin{equation}
	\frac{\partial (\rho \mathbf{v})}{\partial t} + \nabla \cdot \left( \mathbf{v} \rho \mathbf{v} - \mathbf{BB} \right) + \nabla p_\mathrm{tot} = 0,
\end{equation}
\begin{equation}
	\frac{\partial \mathbf{B}}{\partial t} + \nabla \cdot \left( \mathbf{vB} - \mathbf{Bv} \right) = - \nabla \times (\eta \mathbf{J}),
\end{equation}
\begin{equation}
	\frac{\partial e}{\partial t} + \nabla \cdot \left( \mathbf{v}e- \mathbf{BB} \cdot \mathbf{v} + \mathbf{v} p_\mathrm{tot} \right) = \nabla \cdot \left( \mathbf{B} \times \eta \mathbf{J} \right),
\end{equation}
where $e$, $\rho$, $\mathbf{v}$, $\mathbf{B}$ are the total energy density, mass density, velocity, and magnetic field, $p = (\gamma - 1)(e - \rho \mathbf{v}^2 / 2 - {B}^2/2)$ is the thermal pressure, $p_\mathrm{tot} =p + {B}^2/2$ is the total pressure, $\mathbf{J} =\nabla \times \mathbf{B}$ is the electric current density; $\eta$ is the electrical resistivity, and $\gamma$ is the ratio of specific heats. As in this study we are not interested in dissipative processes, we take $\gamma=5/3$, and $\eta = 0$.

The physical quantities were normalised with the use of the following constants: the lengths are normalised to
$L_\mathrm{N}=1$~Mm, magnetic fields to $B_\mathrm{N}=20$~G and densities to $\rho_\mathrm{N}=1.67 \times 10^{-15}$~g~cm$^{-3}$.  The mass
density normalisation corresponds to the electron concentration $n_\mathrm{N} = 10^9$~cm$^{-3}$. The normalising speed
was calculated in the units of $v_\mathrm{N} = {B_\mathrm{N}}/\sqrt{4 \pi \rho_N} = 1,380$~km s$^{-1}$ that is the Alfv\'en speed
corresponding to the values of $B_\mathrm{N}$ and $\rho_\mathrm{N}$, the normalising time was set to $t_\mathrm{N}=L_\mathrm{N}/v_\mathrm{N}=0.7246$~s,  the normalisation of the radial derivative of the azimuthal velocity was set to $\Omega_\mathrm{N}=v_\mathrm{N}/L_\mathrm{N}=1.38$~s$^{-1}$. The values used for the normalisation are typical for the solar corona; their use justifies our intention to analyse nonlinear effects in the coronal plasma. Further in the text we imply normalised units if it is not stated otherwise.

\subsection{Numerical setup}
\begin{figure}[pth]
	\begin{center}
		\includegraphics[width=8.0cm]{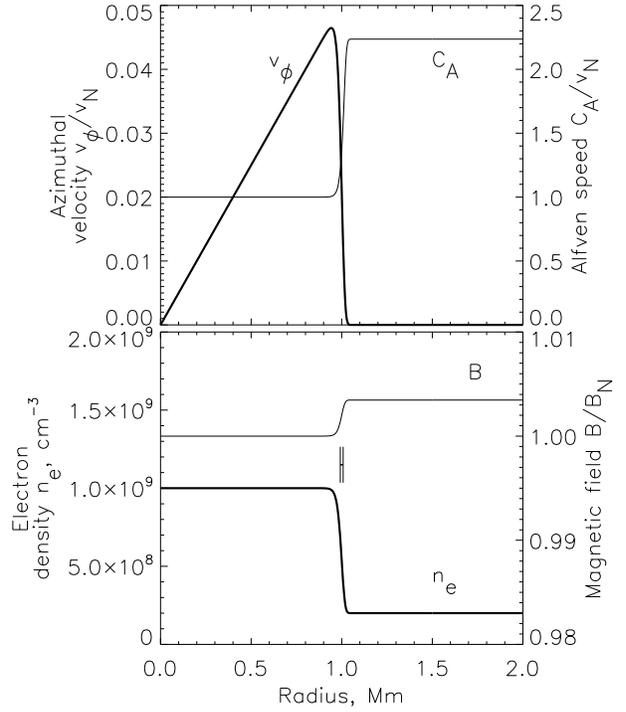} 
	\end{center}
	\caption{Top: Radial profiles of the normalised initial azimuthal velocity $v_\varphi(r)$  and equilibrium Alfv\'en speed $C_\mathrm{A}(r)$ in the simulated magnetic flux tube. Bottom: Radial profiles of the normalised equilibrium electron density $n_e(r)$ and equilibrium magnetic field $B_{z}(r)$. The line segment at the point $(1.0, 10^9)$ in the bottom panel indicates the scale of a single pixel used during the simulation.}
	\label{density_profile}
\end{figure}

We used a cylindrical frame of reference, and considered a straight magnetic tube directed along the $z$-axis of the computational box. The equilibrium magnetic field is parallel to the tube axis. The equilibrium concentration of electrons, $n_e(r)$ was set using the generalised symmetric Epstein function \citep{2003A&A...409..325C}, and  %
the equilibrium parallel magnetic field $B_z(r)$ was set to equalise the equilibrium total pressure everywhere in the computational box,
\begin{equation}
	\left\{ \begin{aligned}
            n_\mathrm{e}(r) & = n_{\infty}+ \frac{(n_0-n_{\infty})} { \cosh^2 \left[ \left(r/R_0\right)^\alpha \right] } \\
	    B_z(r)  & = B_0 \left\{ 1 + \frac{4 T(n_0 - n_{\infty})}{B_0^2}  \left( 1-\frac{1}{\cosh^2 
	              \left[ \left(r/R_0\right)^\alpha \right]} \right) \right\}^{1/2}, \\
    \end{aligned} \right. 
\end{equation}
where $R_0=1$ is the tube radius, $n_{\infty}=0.2$, $n_0=1$, $B_0=1$, and the temperature $T$ is constant throughout the whole volume. In physical units these values correspond to $R_0=1$~Mm, $n_{\infty}=2\times10^8$~cm$^{-3}$, $n_0=10^9$~cm$^{-3}$ and $B_0=20$~G, i.e. typical for the corona. The parameter $\alpha$ controls the tube boundary steepness. We used $\alpha=36$ that results in a rather sharp boundary. The radial profiles of the electron density $n_e(r)$ and magnetic field $B_{z}(r)$ are depicted in Figure~\ref{density_profile}. 
 
All the equilibrium values are constant along the $z$-axis, since in this study we consider neither gravity nor transition region.

The torsional Alfv\'en wave is driven at the bottom wall $z=z_\mathrm{min}=0$ of the equilibrium magnetic flux tube by the following perturbations,
\begin{equation}
  \begin{aligned}
	  B_{\varphi}(r) & =   J_\mathrm{M} r \frac{\sin (\omega t )}{ \cosh^2 \left[
	  \left(r/R_0\right)^\alpha \right] }, \\
	  v_{\varphi}(r) & =   \Omega_\mathrm{M} r \frac{\sin (\omega t)}{ \cosh^2
	  \left[ \left(r/R_0\right)^\alpha \right] }, 
  \end{aligned}
  \label{perturb}
\end{equation}
where $J_\mathrm{M}$ is the amplitude of the radial derivative of the azimuthal component of the magnetic field, $\Omega_\mathrm{M} = - J_\mathrm{M}/\sqrt{4 \pi \rho(r)}$ is the amplitude of the radial derivative of the azimuthal component of the velocity, and the factor $\cosh^{-2} \left[ \left(r/R_0\right)^\alpha \right]$ is introduced to avoid the excitation of the torsional wave outside the tube.  The frequency $\omega$ is set to $\omega=\omega_\mathrm{A}=2\pi/\lambda C_\mathrm{A}$, where $\lambda$ is a wavelength. The top boundary is set to keeps zero gradients of the variables (the \verb{cont{  flag in MPI-AMRVAC). In this study we consider the waves before they reach the top boundary, and hence the specific choice of the top boundary condition is not important.

\subsection{Parameters of numerical experiments}\label{other_parameters}

We analysed dynamics of the torsional Alfv\'en wave, controlling the following parameters: the plasma temperature $T$ (and hence, for the fixed value of $B_0$, on the plasma parameter $\beta$ that is the ratio of the gas and magnetic pressures), amplitude of the derivative of the azimuthal velocity perturbation, $\Omega_\mathrm{M}$, and wavelength $\lambda$. The specific values of these parameters in various numerical runs are shown in Table~\ref{table1}.
\begin{table}
	\centering
	\caption{Parameters of the simulations: the plasma temperature $T$,  wavelength $\lambda$, the
		initial amplitude of the radial derivative of the azimuthal velocity perturbation $\Omega_\mathrm{M}$
	    (normalised), and the plasma parameter $\beta$.}
	\begin{tabular}{l|cccc}
           Title           & $T$,~MK  & $\lambda$,~Mm & $\Omega_\mathrm{M}$ & Plasma $\beta$ \\
	   \hline
	   \hline
	   \textbf{setup1}  & 0.2      & 5.0         & 0.05   & 0.0035\\
           \textbf{setup2}  & 0.2      & 10.0        & 0.05   & 0.0035  \\
           \textbf{setup3}  & 1.0      & 2.0         & 0.05   & 0.0173  \\
           \textbf{setup4}  & 1.0      & 5.0         & 0.05   & 0.0173  \\
           \textbf{setup5}  & 1.0      & 10.0        & 0.05   & 0.0173 \\
	   \textbf{setup6}  & 1.0      & 10.0        & 0.075  & 0.0173 \\
           \textbf{setup7}  & 1.0      & 10.0        & 0.10   & 0.0173  \\
           \textbf{setup8}  & 1.0      & 10.0        & 0.15   & 0.0173 \\
           \textbf{setup9}  & 1.0      & 10.0        & 0.20   & 0.0173 \\
           \textbf{setup10} & 5.0      & 10.0        & 0.05   & 0.087 \\
	\end{tabular}
	
	\label{table1}
\end{table}

In all these setups the relative variation of the magnetic field is small, and the plasma-$\beta$ is less than unity everywhere.

The simulations were performed in a cylindrical frame of reference,  in either a 3D $(r, \varphi, z)$ or 2D $(r, z)$ computational boxes. We used a 3D computational box for the investigation of the general picture of the wave propagation, and high-resolution 2D computational boxes were used for the analysis of wave-steepening, radial profiles and generation of sausage waves. 

In the 3D case the box size was $128 \times 64 \times 256$ grid points, which corresponds to the physical volume $[0, 2]$~Mm in $r$, $[0, 2\pi]$ in $\varphi$, and $[0, 40]$~Mm in $z$. The numerical grid (pixel) size corresponds to $0.0156$~Mm in $r$ (see Figure~\ref{density_profile}) and $0.156$~Mm in $z$, i.e. much less than the characteristic scales of the non-uniformity and wavelength. In order to test the undesirable effect of the grid resolution we carried out a test run with double the spatial resolution. We found that in both the cases the evolution of the torsional wave shows similar behaviour, in general, but the setup with the lower spatial resolution demonstrated faster decay and a smoother wave profile at the tube boundary.

In 2D simulations we used different grids. For the investigation of wave steepening we used the numerical grid $384\times2048$, with $r \in [0,3]$~Mm and $z \in [0,80]$~Mm, respectively. For the investigation of the radial profiles of the perturbations we used the numerical grid $1024\times256$, with $r \in [0,2]$~Mm and $z \in [0,20]$~Mm, respectively, and for the investigation of sausage magnetoacoustic wave generation we used the grid $384\times8192$, with $r \in [0,3]$~Mm and $z \in [0,320]$~Mm. In all the cases the spatial resolution in the \lq\lq important\rq\rq\ dimension was much higher than in the 3D setup.

In MPI-AMRVAC we choose either \verb{HLLC{ (for 3D simulations), or \verb{ssprk54{
(for 2D simulations) discretisation method, with the \verb{vanleer{ slope limiter. The constraint $\nabla B = 0$ is controlled by the \verb{powel{ approach. Since we already analysed the effect of numerical resolution on the results, we use no mesh refinement and set the parameter \verb{mxnest{ to 1.

\section{Results and discussion}\label{results}

\subsection{General picture of torsional wave propagation} 
\label{general}

\begin{figure*}[pth]
	\begin{center}
		\includegraphics[width=16.0cm]{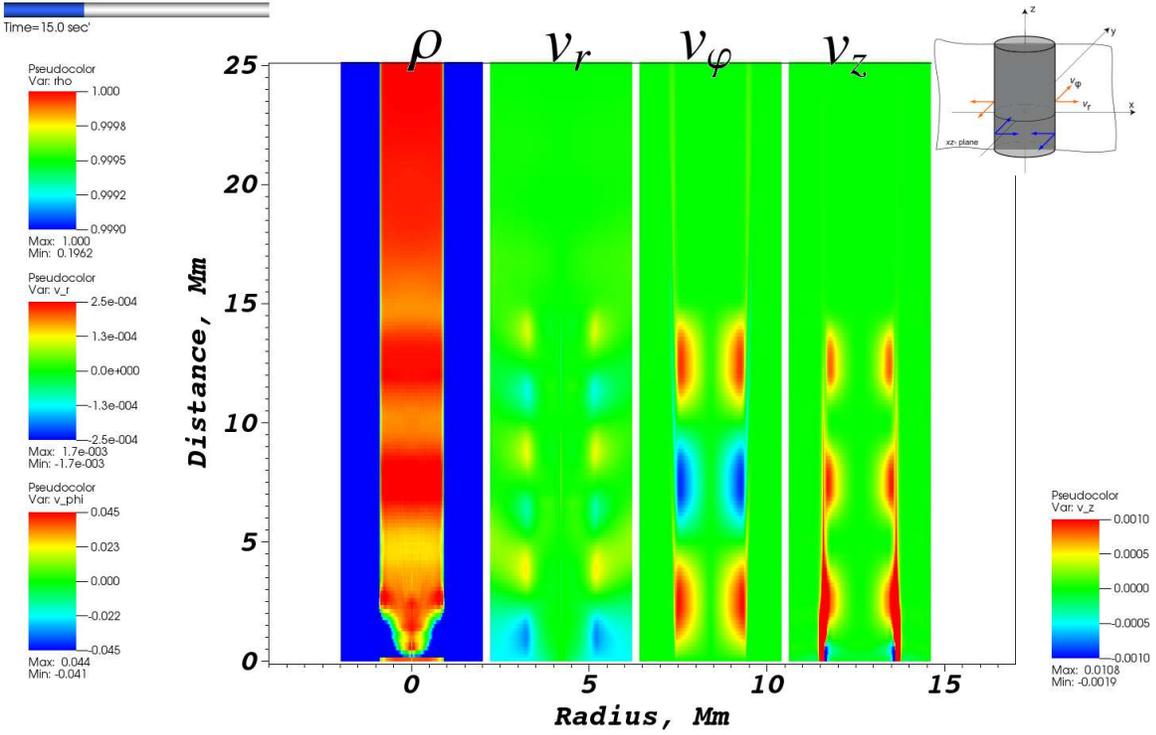}
	\end{center}
	\caption{Torsional Alfv\'en wave propagating along the magnetic flux tube. The panels show the spatial structure
		of the plasma density $\rho$ and velocities $v_r$, $v_\varphi$, $v_z$ (from left to right) in the $xz$
		cross-section through the axis of the flux tube, shown in the inlet. The torsional Alfv\'en wave is
		excited at $z=0$, and propagates in the positive $z$-direction (upwards). The snapshot corresponds to
		setup 5, i.e. $T=1$~MK, $\lambda=10$~Mm, $\Omega_\mathrm{M}=0.05$; the time instant is
	$t=15t_\mathrm{N}=10.9$~s. A relevant animation is available in the online version.}
	\label{3d-snapshot}
\end{figure*}

In general, results of the simulation agree with the theory highlighted in {Appendix~\ref{thintube}}. A torsional wave
appears as an alternate perturbation of the azimuthal plasma velocity accompanied by a perturbation of
azimuthal component of the magnetic field, $B_\varphi$. The wave dynamics preserves the axial symmetry.  Snapshots
demonstrating perturbations of various physical quantities in the wave are shown in Figure~\ref{3d-snapshot}. The wave is
seen to propagate along the flux tube, in the positive $z$-direction, at the speed about the Alfv\'en  speed
$C_\mathrm{A}$ in the body of the flux tube. At the time of the snapshots $t_1=15t_\mathrm{N}=10.9$~s, the wave has
propagated the distance $l_1=C_\mathrm{A} t_1 \approx 15\mbox{~Mm} =  1.5\lambda$ from the point of the excitation,
$z=0$.

The local Alfv\'en speed (see Figure~\ref{density_profile}) increases near the tube boundary, in a thin layer where the equilibrium plasma density decreases. This leads to the outrunning of the torsional wave near the tube boundaries, producing a distortion of the wave fronts. It is a clear signature of phase mixing caused by the transverse non-uniformity of the Alfv\'en speed. We refer to the effect as linear phase mixing and will further consider it in Section~\ref{phasemix}.

The detected torsional wave induces flows with two other velocity components, the radial velocity $v_r$ and parallel velocity $v_z$, and also the density perturbations that propagate at the Alfv\'en speed and have double the frequency of the driver and the perturbations of the azimuthal velocity $v_\varphi$ and magnetic field $B_\varphi$. The amplitudes of induced $v_r$ and $v_z$ amounts to square of the amplitude $v_\varphi$ of the initial wave.

In addition, in the bottom part of the computational domain (Figure~\ref{3d-snapshot}) one can see the development of the tube wave propagating at the tube speed $C_\mathrm{T} \approx 0.15$~Mm/s. At the time of the snapshot, the wave has propagated the distance $l_2=C_\mathrm{T} t_1 \approx 1.6$~Mm from the excitation point, which is consistent with Eq.~(\ref{Atube}). The tube wave is excited when the plasma temperature is finite, and hence the tube speed $C_\mathrm{T}$ is greater than zero.

The appearance of the parallel compressive perturbations characterised by $\rho$ and $v_z$ may be attributed to the effect of the nonlinear ponderomotive force associated with the gradients of the perturbed Alfv\'en speed in the parallel direction, and that is in agreement with the theory developed in \citet{VasheghaniFarahani2011}, see also Eq.~(\ref{Asolution}). The excitation of the radial flows that are compressive too, may be attributed to the nonlinear ponderomotive force associated with the gradients of the perturbed Alfv\'en speed in the perpendicular direction \citep{Nakariakov1997}, see also {Appendix~\ref{Aphasemix}}, and to the effect of nonlinear phase mixing, connected with the radial non-uniformity of the torsional wave amplitude.

\subsection{Radial profiles of induced compressive perturbations}

\begin{figure}[pth]
	\begin{center}
		\includegraphics[width=8.0cm]{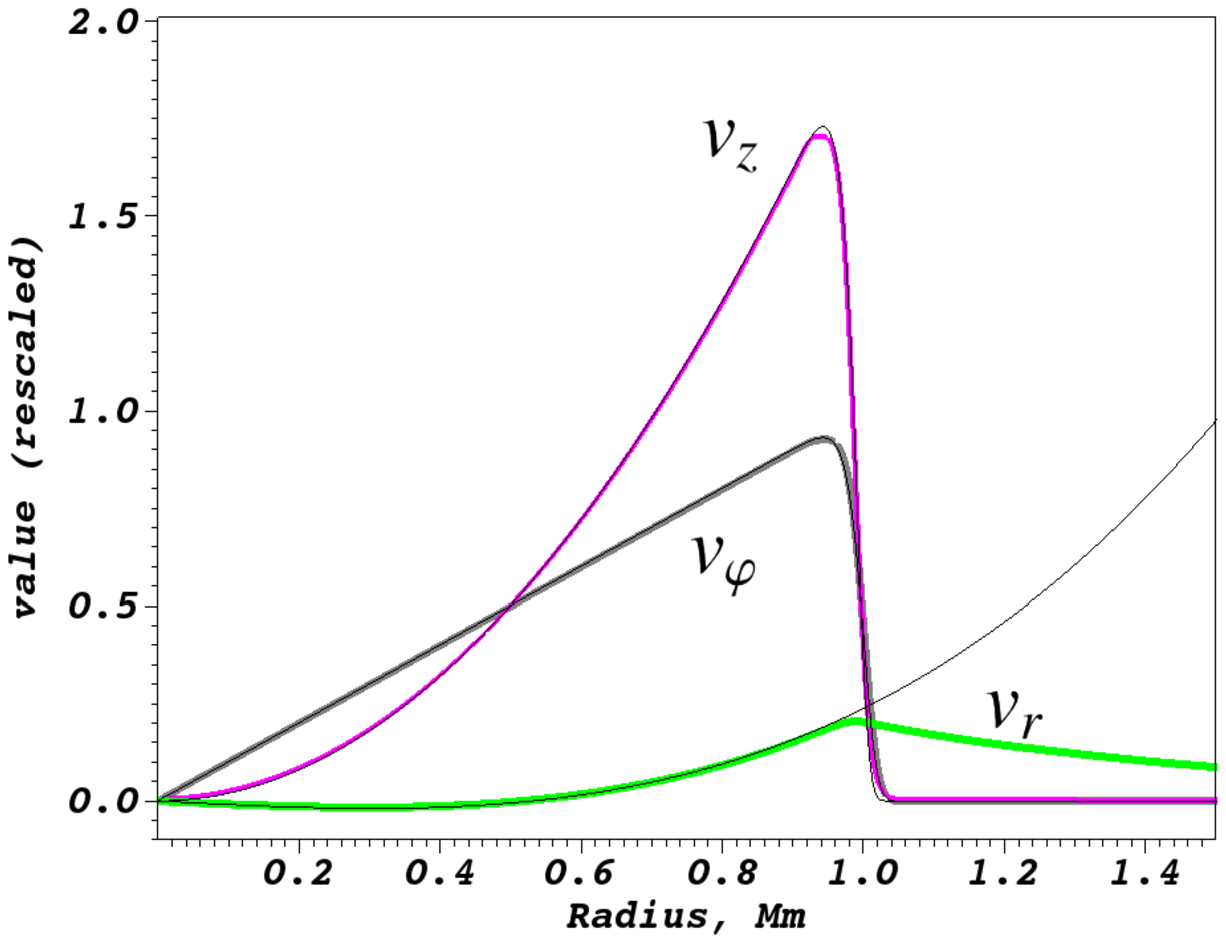}
	\end{center}
	\caption{Radial profiles of $v_r$, $v_\varphi$ and $v_z$ velocities in the wave maximum. Thick curves are the
		results of numerical simulation: the black line shows the azimuthal velocity $v_\varphi$,
		magenta curve shows the parallel velocity $v_z$, and green curve
		the radial velocity $v_r$. The profiles are rescaled for the purpose of visualisation: $v_\varphi$ is divided by
		$\Omega_\mathrm{M}=0.05$, and $v_z$ and $v_r$ are divided by 
		$6.25\times 10^{-4}$.
		 The thin curves indicate the best-fitting analytical dependencies.
	}
        \label{radial_profiles_v}
\end{figure}

The radial structure of the induced parallel flows at the speed $v_z$, obtained in the simulation (see Figure~\ref{radial_profiles_v}) is seen to be different from that described by expressions~(\ref{Avariables}) and (\ref{Aponderomotive}) obtained in the thin flux tube approximation. It clearly shows a parabolic shape, $\propto r^2$. This difference could be attributed to the finite wavelength effects. A similar radial dependence is seen in the radial structure of the perturbed parallel magnetic field $B_z$ (not shown here). 

The simulated radial profile of $v_z$ was found to be well fitted with an almost parabolic dependency,
\begin{equation}
	 v_z(r,z,t) = \frac{\Omega_M^2 r^2}{4 C_\mathrm{A}}
	\frac{ \left( 1 - \cos \left[ 2 \omega (t - z/C_\mathrm{A}) \right] \right) }{ \cosh^4 \left[ (r/R_0)^{36} \right] },
	\label{ponderomotive2n}
\end{equation}
determined empirically, by a guess.

The radial dependence in the denominator is artificially used to reproduce the disappearance of the perturbation at the flux tube boundary.
This parabolic dependence is different from the solution for $u_\mathrm{p}$, given by Eq.~(\ref{Aponderomotive}) that is independent of the radial coordinate $r$. 
We need to emphasise that the dependence given by Eq.~(\ref{ponderomotive2n}) is fully empirical, and is not based on any theoretical result. It is shown here for its possible usefulness for forward modelling of the observational manifestation of this effect, when it would be convenient to have a single functional expression for the radial structure of the induced parallel flows. The perturbations of the density $\rho$ and pressure $p$ do not show dependence on the radial coordinate, and have a constant amplitude across the flux tube.

The radial profiles of $v_r$ and $B_r$ are different from the other variables. They have half the wavelength of the mother torsional wave (given by the perturbations of $v_\varphi$ and $B_\varphi$), and are shifted 
by $+\pi/4$ relative to the mother torsional wave. The radial structure of $v_r$ can be expressed using the odd terms of the Taylor expansion (see Figure~\ref{radial_profiles_v}), beginning with the linear one that is prescribed by the boundary condition at the loop axis 
\citep[see also the discussion in][]{1996PhPl....3...10Z}. 
In the performed numerical run, we found that the radial dependence of the radial velocity inside the flux tube could be best-fitted by the expression $v_r(r) = 0.33r(r^2 - 0.28)$.  
We note that the radial velocity has relatively large values outside the tube's boundary, whereas the azimuthal and parallel velocities do not penetrate into the external medium. 

\subsection{Parallel spatial structure of induced compressive perturbations}

\begin{figure}[pth]
	\begin{center}
		\includegraphics[width=8.0cm]{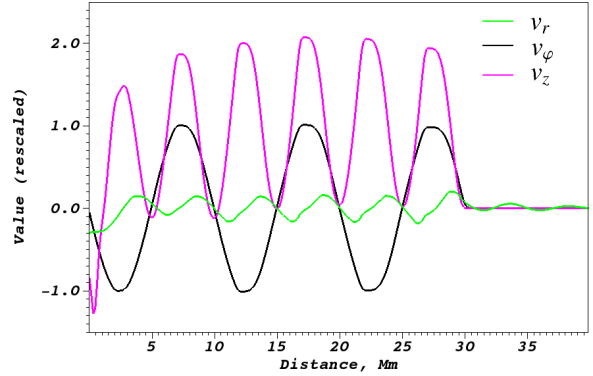}
	\end{center}
	\caption{Spatial profiles of  $v_r$, $v_\varphi$, $v_z$ velocities measured along the $z$-axis at $r_1=0.75$~Mm at
		the time instant $t=30~t_N$. The black curve shows the azimuthal velocity $v_\varphi$, the
		magenta curve the parallel velocity $v_z$, and the green curve the
		radial velocity $v_r$.  The profiles are rescaled for the purpose of visualisation: $v_\varphi$
		is divided by $\Omega_\mathrm{M} r_1=
		3.75\times 10^{-2}$, and $v_r$ and $v_z$ are divided by
		$3.52\times 10^{-4}$. 
	}
	\label{spatial_profiles}
\end{figure}

Parallel spatial profiles of the velocities $v_r$, $v_\varphi$, $v_z$ measured along the $z$-axis at $r_1=0.75$~Mm are shown in Figure~\ref{spatial_profiles}. The profiles are rescaled for the purpose of visualization: $v_\varphi$ is divided by $\Omega_M r_1 = 3.75 \times 10^{-2}$, and $v_r$ and $v_z$ are divided by $\Omega_M^2 r_1^2/ 4C_A=3.52 \times 10^{-4}$. This rescaling corresponds to omitting of amplitudes and radial dependencies of the variables in Eqs.~(\ref{perturb}) and (\ref{ponderomotive2n}) (and with the caveat in Eqs.~(\ref{Atube}) and (\ref{Aponderomotive})).  These profiles are consistent with the theory described in Appendix~\ref{thintube}: the wavelength of the induced parallel flows is a half of the wavelength of the torsional wave, the induced parallel flows have a positive average value, and the amplitude corresponds to one given by Eq.~(\ref{Asolution}), i.e. the amplitude of induced parallel velocities is proportional to squared amplitude of the azimuthal perturbation. 

The positive average value of the parallel velocity means a field-aligned bulk plasma flow \citep[e.g., see][]{1998JGR...10323677O}, which can be referred to as the \lq\lq Alfv\'enic wind\rq\rq. The wind is produced by the ponderomotive force. However, the wind is canceled on the arrival of the slow, tube wave $u_\mathrm{T}$. A similar effect has been studied in detail in the case of plane Alfv\'en waves by \citet{McLaughlin2011}. 

In addition, we see small amplitude perturbations of $v_r$. The amplitude of this perturbation is approximately 5 times smaller than the amplitude of parallel motions, and the speed of propagation in the $z$ direction is higher than the speed of the torsional wave. These perturbations correspond to the third kind of the induced compressive perturbations, the fast magnetoacoustic sausage wave \citep[e.g.][]{1983SoPh...88..179E, 2012ApJ...761..134N, 2016ApJ...833...51Y}. In the low-$\beta$ plasma considered here, this mode is characterised by compressive, mainly radial flows, and propagates at the speed that lies between the Alfv\'en speed inside and outside the tube. Positive and negative half-periods of these perturbations are symmetric. Hence, in contrast with the induced parallel flows, they do not constitute any bulk flow of the plasma.

\subsection{Efficiency of the generation of compressive perturbations}

To study whether the efficiency of the nonlinear generation of compressive perturbations depends on various parameters of the plasma, such as the temperature $T$, wavelength $\lambda$ and the tube's radius, i.e. to check the validity of the approximation given by Eq.~(\ref{Aponderomotive}), we perform a parametric study summarised in Table~\ref{table1}. 

We analysed amplitudes of the parallel velocity perturbations, $v_z$, in the leading cycle of the induced wave, at the distance of 10--15~Mm from the origin in Figure~\ref{3d-snapshot}. According to Eq.~(\ref{Aponderomotive}), the behaviour of the induced ponderomotive wave depends neither on the temperature $T$, nor on the wavelength $\lambda$. For the amplitude of the ponderomotive wave,  numerical setups with different plasma temperatures $T$ and wavelength $\lambda$ were found to show good agreement with Eq.~(\ref{Aponderomotive}) near the boundary of the flux tube. In addition, we saw the tube wave propagating at the tube speed, which is consistent with Eq.~(\ref{Atube}), again, near the boundary. 

\subsection{Nonlinear wave steepening}

\begin{figure}[pth]
	\begin{center}
		\includegraphics[width=8.0cm]{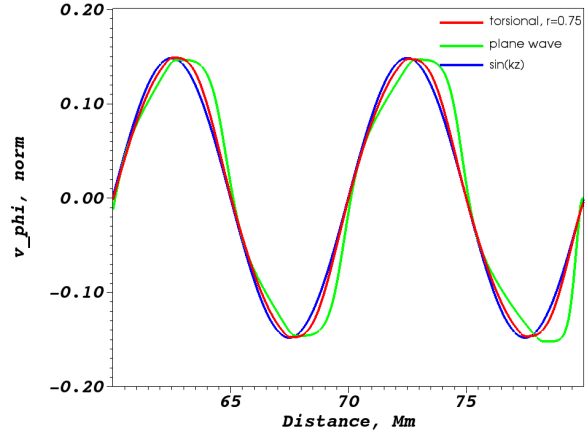}  %
	\end{center}
	\caption{Comparison of snapshots of the azimuthal velocity $v_\varphi$ in a torsional wave (red) and
perpendicular velocity in a plane Alfv\'en wave (green) of the same relative
amplitude. The blue line shows a harmonic function that corresponds to the linear case. Both
waves have propagated the distance of $8\lambda = 80$~Mm from the point of the excitation. The amplitude of the
torsional wave is measured at $0.75$~Mm from the tube axis.}
	\label{torsional_vs_plane}
\end{figure}

The ponderomotive excitation of compressive flows by an Alfv\'en wave, leads to the modification of the local Alfv\'en speed, which, in turn, affects the Alfv\'en wave itself. This chain of events is considered as nonlinear self-interaction of the Alfv\'en wave, which leads to the wave steepening \citep[e.g.][]{1974PhFl...17.2215C}.
In the long wavelength limit, steepening of a torsional wave due to nonlinear self-interaction has been considered by \citet{VasheghaniFarahani2012}. In that study it was found that the torsional wave steeping occurs at the rate that is lower than in the case of plane Alfv\'en waves. The difference between the evolution of the torsional and plane waves disappears in the case of the cold ($\beta = 0$) plasma. In our work we consider this effect on torsional waves of finite wavelength. For comparison, we modelled nonlinear evolution of a plane Alfv\'en wave numerically in a 2D Cartesian geometry, and compared it with the results obtained for a torsional wave in the cylindrical geometry. Snapshots of the parallel, along the equilibrium field, structure of the steepened torsional and plane waves of the same amplitudes and wavelengths are shown in Figure~\ref{torsional_vs_plane}. 

The spatial profiles of both torsional and plane Alfv\'en waves clearly show the steepening: in the extremes, the perturbations in both the waves overtake the harmonic dependence. It occurs in both positive and negative extremes, which is a typical signature of the nonlinear evolution of linearly or elliptically polarised Alfv\'en waves \citep[e.g.][]{1974PhFl...17.2215C, VasheghaniFarahani2012}. In the case of plane Alfv\'en waves this effect is seen to be more pronounced, which is consistent with the results obtained by \citet{VasheghaniFarahani2012}. But, in contrast to the long-wavelength limit, the decrease in the nonlinear steepening of the finite-wavelength torsional waves occurs even in the case of low-$\beta$. It can be attributed to the presence of other sinks of energy, such as the excitation of the sausage mode. 

\subsection{Excitation of a sausage wave}

The radial velocities $v_r$, that were nonlinearly generated by the torsional wave, have an axisymmetric structure, i.e. are independent of the azimuthal coordinate. In a low-$\beta$ plasma, this spatial structure is similar to the structure of sausage (\lq\lq $m=0$\rq\rq) magnetoacoustic modes of the magnetic flux tube, which is an essentially compressive perturbation of the tube. Sausage modes are collective perturbations of the flux tube, and fill in the whole flux tube. Their specific properties, such as the radial structure and dispersion, are prescribed by the specific radial profile of the fast speed \citep[e.g.][]{2012ApJ...761..134N, 2016ApJ...833...51Y}. Consideration of this effect is out of scope of this paper. 

In a low-$\beta$ plasma, a fast sausage mode propagates at the phase speed $v_\mathrm{saus}$ that is lower than Alfv\'en speed of external media $C_\mathrm{A}^\mathrm{ext} = C_\mathrm{A}(r\to\infty)$, but higher than the Alfv\'en speed of plasma near the tube's axis, $C_\mathrm{A}$ \citep{1983SoPh...88..179E}. As fast sausage modes are highly dispersive, a broadband pulse develops in a sausage wave train. For example, \citet{Shestov2015} numerically studied the dispersive evolution of fast sausage wave trains guided by a magnetic flux tube, and confirmed the formation of wave trains with pronounced modulation of the instant period and amplitude. Also it was found that the propagation speed is higher than the local Alfv\'en speed in the flux tube, and hence of the torsional wave. Thus, the nonlinearly induced fast sausage wave should propagate faster than the mother torsional wave.

\begin{figure}[pth]
	\begin{center}
		\includegraphics[width=8.0cm]{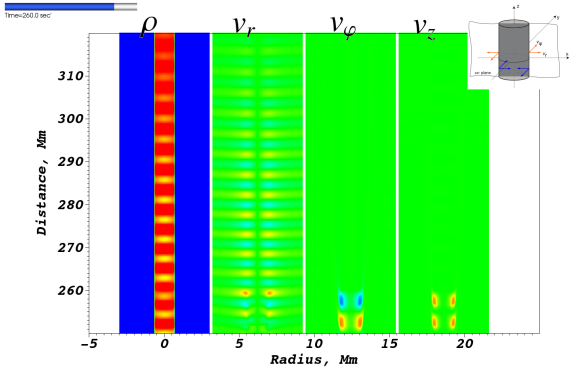}
	\end{center}
	\caption{Spatial structure of perturbations in the vicinity of the
		leading front of the torsional wave, at the distance of
		$26\lambda$ from the excitation point, at the time instant
		$t=260 t_\mathrm{N}=188$~s.		}
	\label{sausage_wave}
\end{figure}

Figure~\ref{sausage_wave} shows the spatial structure of the incompressive and compressive perturbations in the vicinity of the leading front of the torsional wave, with the clear evidence of the fast sausage wave characterised by the perturbations of the density $\rho$ and radial flows $v_r$ preceding the torsional wave demonstrated by the perturbations of $v_\varphi$ and the ponderomotively induced parallel flows $v_z$. The sausage perturbations are seen to propagate at the speed $v_\mathrm{saus} \approx 1.3C_\mathrm{A}$, which is consistent with the theoretical estimation of \citet{1983SoPh...88..179E}. The amplitude modulation of the sausage wave is also evident. Its amplitude varies in the $z$-direction, and, in particular, has a maximum at $z\approx 280$~Mm at the instant of time of the snapshot shown in Figure~\ref{sausage_wave}.

\subsection{Linear and nonlinear phase mixing}
\label{phasemix}
\begin{figure}[pth]
	\begin{center}
		\includegraphics[width=8.0cm]{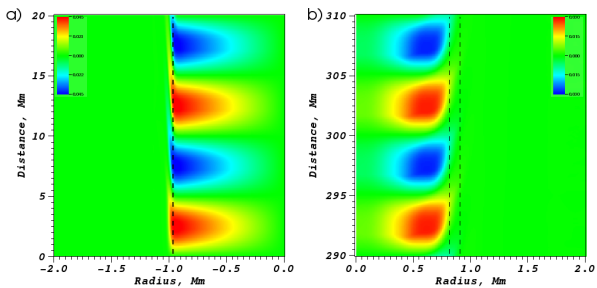}
	\end{center}
	\caption{Comparison of linear and nonlinear phase mixing in a torsional Alfv\'en wave with the relative amplitude 0.05. Left: snapshot of the azimuthal velocity in the torsional wave that travelled 20~Mm from the source. Right: the same but at the distance of 310~Mm from the source. The vertical dashed lines highlight the radial distances discussed in the text. }
	\label{phase-mixing}
\end{figure}

The effect of linear phase mixing occurs near the tube boundary, where the local Alfv\'en speed gradually increases due to the decrease in the plasma density. The efficiency of phase mixing can be estimated using the following kinematic reasoning. The difference $\delta$ between the distance travelled by the torsional perturbation near the boundary and in the body of the tube equals to the product of the difference of the local Alfv\'en speeds at these radial surfaces, $\Delta C_\mathrm{A}$ and the travel time $t$. Thus the travel time required for phase mixing of one wavelength $\delta = \lambda$ is the ratio of the wavelength in the body of the tube and the speed difference $t=\lambda / \Delta C_\mathrm{A}$. The travel time equals to the distance travelled by the wave from the source, divided by the Alfv\'en speed in the body of the tube $t=z/C_\mathrm{A}$. Thus, the distance at which phase mixing reaches one wavelength equals to $z=\lambda C_\mathrm{A} / \Delta C_\mathrm{A} $.
In particular, in the numerical setup shown in Figure~\ref{density_profile}, the local Alfv\'en speed increases near the boundary by about $\Delta C_\mathrm{A} \approx 1.2 C_\mathrm{A}$, hence for the wavelength of 10~Mm the phase mixing reaches the wavelength at the distance of about 8.3~Mm from the wave source. This value is consistent with the deformation of the torsional wave front seen in the left panel of Figure~\ref{phase-mixing}.

Comparing the shapes of the torsional perturbations taken at different distances from the source in
Figure~\ref{phase-mixing}, we see that at large distances from the source, the perturbation shape experiences some deformation even in the body of the flux tube. More specifically, near the source (left panel),  the wave front is clearly deformed because of the non-uniformity of the local Alfv\'en speed in the region $r>0.95$~Mm. At $r<0.95$~Mm, near the source the wave fronts are symmetric with respect to the parallel coordinate.
On the other hand,  at the larger distance from the source (right panel), the wave front becomes clearly deformed even at the radial distances of $r < 0.8$~Mm. This deformation has the typical signature of phase mixing: the perturbations at larger radial distance from the axis propagate slightly faster. However, at these radial locations the relative change of the local Alfv\'en speed is negligible. Thus, this phase mixing is not connected with the non-uniformity of the local Alfv\'en speed. The wave front deformation should be attributed to the variation of the propagation speed of the torsional wave caused by the variation of its amplitude with the radial coordinate. As the torsional wave amplitude increases in the radial direction, it causes additional radial non-uniformity of the torsional wave speed because of the nonlinear acceleration. It can further enhance phase mixing of torsional waves --- the effect we shall refer to as \lq\lq nonlinear phase mixing\rq\rq. In the right panel the torsional perturbation is absent from the region $r > 0.8$~Mm, because of the numerical dissipation of the very strong phase mixing occurring in this region at much shorter distances from the source. 

Following the reasoning used in the case of linear phase mixing, the distance at which nonlinear phase mixing reaches one wavelength equals to the product of the wavelength and the ratio of the local Alfv\'en speed and the increase in the speed caused by the nonlinearity, $z=\lambda C_\mathrm{A} / \Delta C_\mathrm{A}^\mathrm{nl}$, where $\Delta C_\mathrm{A}^\mathrm{nl}$ is the difference in the local propagation speeds between the nonlinear and linear torsional waves. The increase in the local speed of the torsional wave, $\Delta C_\mathrm{A}^\mathrm{nl}$, is proportional to the product of the local Alfv\'en speed and the square of its relative amplitude (see the second term in the brackets on the right hand side of Eq.~(22) of \citet{VasheghaniFarahani2012}). As in the performed simulation the relative amplitude of the torsional wave amplitude was 0.05, the effect of nonlinear phase mixing should be rather weak. More specifically, in this run the efficiency of nonlinear phase mixing more than a hundred of times ($0.05^2$) weaker than of the linear phase mixing that operates near the boundary. Indeed, according to Figure~\ref{phase-mixing}, at the distance of about 30 wavelengths from the source, phase mixing equals to a fraction of the wavelength.

\section{Conclusions}\label{conclusions}

We performed numerical simulations of axisymmetric finite-amplitude torsional Alfv\'en waves in a field-aligned magnetic flux tube filled in with a low-$\beta$ plasma. The flux tube is straight, untwisted and non-rotating. The flux tube is surrounded by a plasma with the magnetic field directed in the same direction as inside the tube, in the direction of the tube axis. The plasma density inside the flux tube is enhanced, which makes the tube a fast magnetoacoustic waveguide. The plasma parameter $\beta$ is taken to be small everywhere. The torsional perturbations are excited at one footpoint of the tube as alternate periodic azimuthal rotations of the tube. The wavelength of the driven torsional wave is of the same order as the diameter of the flux tube. In the linear regime, the torsional wave is incompressive and consists of the alternate azimuthal flows and the perturbations of the azimuthal component of the magnetic field.

We found that nonlinear evolution of the torsional Alfv\'en wave leads to the excitation of three different kinds of compressive motions that propagate along the axis of the magnetic flux tube: the well-known parallel flow of the plasma at the Alfv\'en speed; another parallel flow, at the tube speed, that is the slow magnetoacoustic wave; and, in addition, the mainly radial axisymmetric perturbations propagating at the speed higher than the Alfv\'en speed inside the flux tube, i.e. faster than the speed of the mother torsional waves. The latter kind of induced waves is the sausage fast magnetoacoustic modes. The sausage mode perturbs the plasma also outside the flux tube, which is consistent with the radial structure of sausage modes. 
Nonlinearly induced flows of all three kinds have double the frequency of the mother torsional wave. 

The nonlinearly induced parallel plasma flow that propagates at the Alfv\'en speed can be considered as the \lq\lq Alfv\'enic wind\rq\rq, as its average over the oscillation period is not zero. The Alfv\'enic wind is absent from the vicinity of the flux tube axis, where the torsional wave amplitude is always zero. Thus, the parallel plasma flow that is nonlinearly induced by a torsional Alfv\'en wave has an annulus shape. The effect of Alfv\'enic wind has been concluded to be a possible mechanism for the acceleration of solar and stellar winds \citep[e.g.][]{1998JGR...10323677O, 2011SSRv..158..339S}, but those estimations were based on the assumption of plane Alfv\'en waves.  Our study demonstrates that the field-aligned compressive flows induced by Alfv\'en waves in the solar and stellar coronae are essentially non-uniform in the transverse (e.g., the horizontal direction in the case of open magnetic configurations in coronal holes). There are always regions that are situated near the axes of the wave-guiding magnetic flux tubes, where the Alfv\'enic winds are zero. It may have important implications for the solar and stellar wind acceleration problems, which require a dedicated study.

The intrinsic radial non-uniformity of the torsional wave amplitude, which is connected with the need to satisfy the zero boundary condition at the axis of the flux tube, leads to the effect of nonlinear phase mixing. This effect is connected with the nonlinear increase in the wave speed at the radial shells where the wave amplitude is higher. It leads to the additional distortion of the torsional wave front and hence the generation of progressively small scales in the radial direction. In particular, this effect should enhance the nonlinear generation of the sausage modes. For example, for a simple, linearly growing in the radial direction radial profile of the torsional wave amplitude, the wave front propagates faster near the flux tube boundary because of the nonlinear effects. Thus, phase mixing occurs even if the radial profile of the Alfv\'en wave is flat, when this effect is absent for linear Alfv\'en waves. Further investigation of this effect would be of interest. 

The induced compressive flows modify the local Alfv\'en speed, causing the self-interaction of torsional Alfv\'en waves. It causes the wave profile steepening that is a signature of the nonlinear cascade along the field. The comparison of the nonlinear steepening of the torsional wave with this effect in a plane Alfv\'en wave of the same amplitude and wavelength showed that the efficiency of the parallel nonlinear cascade in a torsional wave is lower than in a plane Alfv\'en wave. It could be attributed to the presence of additional sinks for the torsional wave energy, for example, the excitation of the sausage wave. The steepening takes place at some radial distance from the flux tube axis, and is strongest where the torsional wave amplitude is highest, e.g. near the flux tube boundary in the case of a linear radial profile of the torsional wave amplitude.

For the linear profile of the torsional wave amplitude, we found that the amplitude of the induced parallel flows has a parabolic radial structure. It is not a surprise, as the amplitude of the parallel motions is proportional to the mother torsional wave amplitude squared: the parallel flows could not be nonlinearly induced near the flux tube axis, where the mother wave amplitude tends to zero. This result is different from the flat radial dependence assumed in the thin flux tube approximation given by expressions (\ref{Avariables}). However, it does not show a problem with the thin flux tube approximation, as there the physical quantities are taken either at the flux tube axis or its boundary. The empirically determined best-fitting dependences of the radial profiles of the velocity vector components could be of interest for forward modelling of torsional waves, e.g. in the further development of the studies of \citet{2016FrASS...3....4V}.

Our study demonstrates the importance of the transverse profile for the evolution of coronal torsional waves. The follow-up work should account important effects of stratification, and the variation of the flux tube area and plasma temperature.
 
\section{Acknowledgements}

The work was partially supported by the Program for Fundamental Research of the Presidium RAS P7 ``Experimental and theoretical investigations of solar system and star planetary system objects. Transitional and explosion process in astrophysics'', by Russian Science Foundation (RSF) grant No. 17--12--01567. SVS acknowledges support from the Belgian Federal Science Policy Office through the ESA --- PRODEX programme (grant No.~4000117262). VMN acknowledges the support by the European Research Council under the \textit{SeismoSun} Research Project No.~321141. The simulations were supported by the Supercomputing Center of Lomonosov Moscow State University \citep{Lomonosov} and at Royal Observatory of Belgium. The results of the simulations were analysed with VisIt software~\citep{visit}. The authors are grateful to Chun Xia for his help with MPI-AMRVAC code.  

\appendix

\section{Weakly nonlinear torsional waves in thin flux tube approximation}
\label{thintube}

We illustrate the excitation of compressive perturbations by the analysis of long-wavelength weakly nonlinear torsional Alfv\'en waves. Consider perturbations of a straight cylindrical magnetic flux tube, using the cylindrical coordinate axis with the axis $z$ coinciding with the flux tube's axis. A torsional Alfv\'en wave consists of alternate twisting azimuthal motions $v_\varphi$ of the plasma, accompanied by the azimuthal components of the magnetic field $B_\varphi$. Since both quantities should vanish on the axis of the magnetic tube, torsional Alfv\'en waves could be considered in terms of the second order thin flux-tube approximation of \citet{1996PhPl....3...10Z}. In this approach, the parallel wavelength of the perturbations is taken to be much larger than the flux tube radius, which allows one to consider only a few lowest order terms in the Taylor expansion with respect to the radial coordinate,
 \begin{equation}
  \label{Avariables}
  \begin{aligned}
    & \rho \approx \tilde{\rho}, \, p \approx \tilde{p}+p_2 r^2, \, v_r \approx Vr, \, v_{\varphi} \approx
    \Omega r, \, v_z \approx u \\
    & B_r \approx B_{r1}r, \, B_{\varphi} \approx Jr, \, B_z \approx \widetilde{B_z} + B_{z2}r^2, 
  \end{aligned}
\end{equation}
where $B_r$, $B_{\varphi}$, and $B_z$ are the radial, azimuthal and longitudinal components of the magnetic field, and $v_r$, $v_{\varphi}$, and $v_z$ are the radial, azimuthal and parallel components of the velocity, respectively; $V$, $\Omega$ and $J$ are the radial derivates of the radial and azimuthal components of the velocity, and the azimuthal field, respectively; $\rho$ is the mass density; and $p$ is the gas pressure. The quantities with the overtilde are the zeroth order terms of the expansions with respect to the radial coordinate. The overtilde will be omitted hereafter. In the derivation of these equations it is assumed that the longitudinal wavelength of the perturbations is much larger than the radius of the flux tube. 

Applying expansion (\ref{Avariables}) to the ideal MHD equations, and omitting terms with higher degrees of $r$, one obtains
\begin{equation}
	\rho \left( \frac{\partial V}{\partial t} + u \frac{\partial V}{\partial z} + V^2 - \Omega^2 \right) + 2p_2 +
	\frac{1}{2\pi}\left(J^2+B_zB_{z2}\right) - \frac{1}{4\pi}B_z\frac{\partial B_{r1}}{\partial z} = 0, 
\end{equation}
\begin{equation}
	\frac{\partial \Omega}{\partial t} + u\frac{\partial \Omega}{\partial z} + 2V\Omega +
	\frac{J}{4\pi\rho}\frac{\partial B_z}{\partial z}-\frac{B_z}{4\pi\rho}\frac{\partial J}{\partial z} = 0,
	\label{Avpi}
\end{equation}
\begin{equation}
  \rho \left( \frac{\partial u}{\partial t} + u\frac{\partial u}{\partial z} \right) + \frac{dp}{dz} = 0,
  \label{Avz}
\end{equation}
\begin{equation}
	\frac {\partial \rho}{\partial t} + \frac{\partial (\rho u)}{\partial z} + 2\rho V = 0,
	\label{Amomentum}
\end{equation}
\begin{equation}
	\frac{\partial B_{r1}}{\partial t} + \frac{\partial (u B_{r1})}{\partial z} - \frac{\partial (V B_z)}{\partial z} = 0,
\end{equation}
\begin{equation}
	\frac{\partial J}{\partial t} + \frac{\partial (uJ)}{\partial z} - \frac{\partial \Omega B_z}{\partial z} + 2VJ - 2\Omega B_{r1} = 0,
	\label{ABphi}
\end{equation}
\begin{equation}
	\frac{\partial B_z}{\partial t} + u\frac{\partial B_z}{\partial t} + 2B_zV=0,
	\label{ABz}
\end{equation}
\begin{equation}
	\left( \frac{\partial }{\partial t} + u \frac{\partial }{\partial z} \right) \frac{p}{\rho^\gamma} = 0,
	\label{ACs}
\end{equation}
\begin{equation}
	2B_{r1} + \frac{\partial B_z}{\partial z} = 0.  
\end{equation}
The effect of the gravitational force is neglected. All considered physical parameters are independent of the azimuthal coordinate, i.e. $\partial/\partial \varphi=0$. In other words, we restrict our attention to the consideration of the axisymmetric perturbations only.

The equations are supplemented by the magnetic flux conservation equation and total pressure balance at the tube boundary,
\begin{equation}
	B_z A = const,
	\label{Aflux}
\end{equation}
\begin{equation}
	p + \frac{B_z^2}{8\pi} - \frac{A}{2\pi} \left[ \rho \left( \frac{\partial V}{\partial t} + u\frac{\partial
	V}{\partial z} + V^2 - \Omega^2 \right) \right. +  \left. \frac{1}{4\pi} \left( J^2 - \frac{1}{4}\left( \frac{\partial
	B_z}{\partial z}\right)^2 + \frac{B_z}{2}\frac{\partial^2 B_z}{\partial z^2}\right) \right] = p_T^\mathrm{ext},
	\label{Apressure}
\end{equation}
where $A=\pi R^2$ is the cross-sectional area of the tube of radius $R$, and $p_T^\mathrm{ext}$ is the external total pressure.
 
The equations are linearised with respect to the equilibrium that is an untwisted and non-rotating flux tube, $\rho_0$, $p_0$, $B_{z0}$, $A_0$ (or $R_0$), and $u_0 = V_0 = J_0 = \Omega_0 = B_{r1}^0 = 0$. All equilibrium quantities are constant.

In the linear regime, torsional motions are decoupled from compressible motions.
Torsional perturbations given by $\Omega$ and $J$ and described by Eqs.~(\ref{Avpi}) and (\ref{ABphi}), 
\begin{equation}
	\frac{\partial \Omega}{\partial t} - \frac{B_{z0}}{4\pi\rho_0} \frac{\partial J}{\partial z} = 0,
\end{equation}
\begin{equation}
	\frac{\partial J}{\partial t} - B_{z0} \frac{\partial \Omega}{\partial z} = 0,
\end{equation}
which are readily combined in the wave equation
\begin{equation}
\label{linAE}
	\frac{\partial^2 J}{\partial t^2} - C^2_\mathrm{A} \frac{\partial^2 J}{\partial z^2} = 0,
\end{equation}
where $\omega/k=C_\mathrm{A}$ and $C_\mathrm{A} = B_{z0}/\sqrt{4\pi\rho_0}$ is the Alfv\'en speed. A harmonic torsional wave is $J=J_\mathrm{M} \cos(\omega t -
k z)$ and $\Omega = \Omega_\mathrm{M} \cos(\omega t- k z)$, where the constant amplitudes $\Omega_\mathrm{M} = - J_\mathrm{M}/\sqrt{4\pi\rho_0}$. 

Compressive perturbations given by $u$, $V$, $\rho$, $p$, $B_z$, $A$, and linked by Eqs.~(\ref{Avz}), (\ref{Amomentum}), (\ref{ABz}), (\ref{ACs}), (\ref{Aflux}) and (\ref{Apressure}). Excluding all variables but $u$, 
one can readily obtain the wave equation
\begin{equation}
\label{Awave_eq0}
\frac{\partial^2 u}{\partial t^2} - C_\mathrm{T}^2 \frac{\partial^2 u}{\partial z^2} = 0
\end{equation}
where $C_\mathrm{T} = C_\mathrm{A} C_\mathrm{S} / (C_\mathrm{A}^2 + C_\mathrm{S}^2)^{1/2}$ is the tube speed, and $C_\mathrm{S}=\sqrt{\gamma p_0/\rho_0}$ is the sound speed. Equation~(\ref{Awave_eq0}) has a propagating wave solution $u=U_m \cos (\omega t - k z)$, where $U_m$ is a constant amplitude, and $\omega/k=C_\mathrm{T}$.

Consider nonlinear interaction of torsional and compressible waves of finite amplitudes. 
Taking into account the quadratically nonlinear term containing the torsional variables in the derivation of Eq.~(\ref{Awave_eq0}), we obtain the inhomogeneous wave equation
\begin{equation}
	\frac{\partial^2 u}{\partial t^2} - C_\mathrm{T}^2 \frac{\partial^2 u}{\partial z^2} = - \frac{R_0^2 C_\mathrm{T}^2}{4\pi \rho_0
	C_\mathrm{S}^2} \frac{\partial}{\partial t} \left( J \frac{\partial J}{\partial z} \right).
	\label{Awave_eq2}
\end{equation}
The right hand side term of Eq.~(\ref{Awave_eq2}) describes the ponderomotive force. The solution of the equation is a sum of solutions of the homogeneous and inhomogeneous equations, i.e. the longitudinal, or \lq\lq tube\rq\rq\ wave propagating at the tube speed, $C_\mathrm{T}$, and the induced, \lq\lq ponderomotive\rq\rq\  wave propagating at $C_\mathrm{A}$. The ponderomotive wave constitutes the Alfv\'enic wind. 

If the torsional wave is driven by a harmonic oscillation with the frequency
$\omega$ at a certain location $z=0$, the solution of the equation is:
\begin{equation}
	u = \left\{ 
		\begin{aligned}
		   u_\mathrm{T}+ u_\mathrm{p}, \quad  & 0    < z < C_\mathrm{T}t, \\
		   u_\mathrm{p},      \quad  & C_\mathrm{T}t  < z < C_\mathrm{A}t, \\
		   0,        \quad  & z >  C_\mathrm{A}t ,
		\end{aligned}
	\right.
	\label{Asolution}
\end{equation}
where the tube wave $u_\mathrm{T}(z,t)$ is
\begin{equation}
	u_\mathrm{T} = \frac{R_0^2 J_\mathrm{M}^2}{16\pi\rho_0 C_\mathrm{A}} \left\{ \cos \left[ 2 \omega (t - z/C_\mathrm{T}) \right] - 1 \right\} = \
	{ \frac{R_0^2 \Omega_\mathrm{M}^2}{4 C_\mathrm{A}} \left\{ \cos \left[ 2 \omega (t - z/C_\mathrm{T}) \right] - 1 \right\} },
	\label{Atube}
\end{equation}
and the ponderomotive wave $u_\mathrm{p}(z,t)$ is
\begin{equation}
	u_\mathrm{p} = \frac{R_0^2 J_\mathrm{M}^2}{16\pi\rho_0 C_\mathrm{A}} \left\{1 - \cos \left[ 2 \omega (t - z/C_\mathrm{A}) \right] \right\} = \
	{ \frac{R_0^2 \Omega_\mathrm{M}^2}{4 C_\mathrm{A}} \left\{1 - \cos \left[ 2 \omega (t - z/C_\mathrm{A}) \right] \right\} } .
	\label{Aponderomotive}
\end{equation}
Equation~(\ref{Asolution}) indicates that in the weakly nonlinear case, torsional Alfv\'en wave induces parallel flows that consist of two motions, the tube wave $u_\mathrm{T}$ and the ponderomotive wave $u_\mathrm{p}$, both of which have the amplitude $R_0^2 \Omega_\mathrm{M}^2 / 4 C_\mathrm{A}$ and the frequency that is double the driving frequency. These parallel flows are accompanied by the perturbations of the plasma density $\rho$. Hence the nonlinearly induced perturbations are compressive. 

However, this formalism does not allow one to take into account the effect of nonlinear phase mixing, connected with the increasing non-uniformity of the torsional wave fronts across the equilibrium magnetic field.

\section{Weakly nonlinear effects associated with Alfv\'en wave phase mixing}
\label{Aphasemix}

Consider a plane Alfv\'en wave in a plasma with a 1D non-uniformity of the Alfv\'en speed across the field. 
For simplicity the plasma is taken to be of zero-$\beta$. In this consideration we follow the formalism developed in \citep{Botha2000}. Let the equilibrium magnetic field of the strength $B_0$ be directed along the $z$-axis.  The equilibrium density of the plasma, $\rho_0$, varies in the $x$-direction. A linearly polarised Alfv\'en wave is characterised by the perturbations of $B_y$ and $V_y$. 

In the weakly nonlinear case we can restrict our attention to the quadratically nonlinear terms only. In this case, the nonlinearly induced flows are described by the equations
\begin{equation}
\label{Bvz}
\frac{\partial^2 V_z}{\partial t^2}  =  - \frac{1}{\rho_0}\left[ \frac{\partial}{\partial t} \left(B_y \frac{\partial B_y}{\partial z}\right)\right],
\end{equation}
\begin{equation}
\label{Avx}
\frac{\partial^2 V_x}{\partial t^2} - C_\mathrm{A}^2(x) \left( \frac{\partial^2 V_x}{\partial x^2} + \frac{\partial^2 V_x}{\partial z^2}\right) = - \frac{1}{\rho_0}\left[ \frac{\partial}{\partial t} \left(B_y \frac{\partial B_y}{\partial x}\right)\right].
\end{equation}
Both these flows are essentially compressive, as both cause the density perturbation,
\begin{equation}
\label{Aden}
\rho = - \int \left[\frac{\partial}{\partial x} \left(\rho_0 V_x\right) + \frac{\partial}{\partial z} \left(\rho_0 V_z\right) \right]\,dt.
\end{equation}

Equation~(\ref{Bvz}) is similar to Equation~(\ref{Awave_eq2}), in the zero-$\beta$ limit. It describes the nonlinear excitation of the parallel plasma flows by the Alfv\'en wave, the ponderomotive wave  or  the Alfv\'enic wind. It is clear that the induced parallel flows have the highest speed at the magnetic field lines where the amplitude of the Alfv\'en wave is the highest. 

Equation~(\ref{Avx}) that describes the nonlinearly induced perpendicular flows is essentially different. Its left hand side describes freely propagating fast magnetoacoustic waves that are subject to refraction connected with the non-uniformity of $C_\mathrm{A}(x)$. In particular, in a magnetic flux tube with the enhanced plasma density, this effect leads to the appearance of fast magnetoacoustic modes, for example, sausage modes \citep[e.g.][]{2012ApJ...761..134N}. The nonlinear excitation of the perpendicular compressive flows occurs when the Alfv\'en wave is non-uniform in the perpendicular direction. Because of Alfv\'en wave phase mixing, the right hand side term experiences continuous growth, magnifying the effect of the nonlinear excitation. 


\end{document}